\documentclass[aps, prb, singlecolumn, eqsecnum, showpacs,a4paper]{revtex4}

\usepackage{epsfig}
\usepackage{amsmath}
\usepackage{amssymb}
\usepackage{mathrsfs}
\usepackage{mathtools}
\usepackage{comment}
\usepackage{color}
\usepackage{braket}
\usepackage{nicefrac}
\usepackage{yfonts}
\usepackage{enumitem}
\newcommand{\be}{\begin{equation}}
\newcommand{\ee}{\end{equation}}
\newcommand{\ba}{\begin{aligned}}
\newcommand{\ea}{\end{aligned}}
\newcommand{\tr}[2]{\mathrm{Tr}_{#1}\Bigl[#2\Bigr]}

\newcommand{\bea}{\begin{eqnarray}}
\newcommand{\eea}{\end{eqnarray}}
\def\nn{\nonumber\\}
\def\fr#1{(\ref{#1})}
\def\L{{\cal L}}
\def\M{{\cal M}}

\newcommand{\gotha}{\text{\textgoth{a}}}
\newcommand{\gothb}{\text{\textgoth{b}}}

\begin{document}
\title{Stationary behaviour of observables after a quantum quench in
the spin-1/2 Heisenberg XXZ chain} 
\author{Maurizio Fagotti and Fabian H.L. Essler}
\affiliation{\mbox{The Rudolf Peierls Centre for Theoretical Physics,
    Oxford University, Oxford, OX1 3NP, United Kingdom}}
\begin{abstract}
We consider a quantum quench in the spin-1/2 Heisenberg XXZ chain. 
At late times after the quench it is believed that the expectation
values of local operators approach time-independent values, that are
described by a generalized Gibbs ensemble. Employing a quantum
transfer matrix approach we show how to determine short-range
correlation functions in such generalized Gibbs ensembles for a class
of initial states.
\end{abstract}
\maketitle

\section{Introduction}
Nonequilibrium dynamics in closed quantum systems, and in particular
quantum quenches, have attracted much experimental
\cite{uc,kww-06,tc-07,tetal-11,cetal-12,getal-11}
and theoretical 
\cite{rev,gg,rdo-08,cc-07,caz-06,bs-08,KE:2008,rsms-08,mwnm-09,BPGDA:2009,fm-10,bkl-10,gce-10,2010_Mossel_NJP_12,CEF,free,rf-11,cic-11,KCI:2011,P:2011,2012_Mossel_NJP_14,2012_Caux_PRL_109,2012_Foini_JSTAT_P09011,2007_Gritsev_PRL_99,2009_Moeckel_AP_324,2011_Pozsgay_JSTAT_P01011,2011_Mitra_PRL_107,2012_Karrasch_PRL_109,2011_Cassidy_PRL_106,gurarie,DT:2012,FE:2013,CE:2013,M:2013}
attention in recent years. There is a growing consensus that
integrable models exhibit important differences in behaviour as
compared to non-integrable ones\cite{nonint}. In particular, by now there is ample
evidence that the stationary state after a quantum quench in an
integrable theory is described by a generalized Gibbs ensemble (GGE)\cite{gg}
with density matrix 
\be
\rho_{\rm GGE}=\frac{1}{Z_{\rm GGE}}\exp\left(-\sum_{l=1}\lambda_lH^{(l)}\right).
\label{rhoGGE}
\ee
Here $H^{(1)}$ is the Hamiltonian and $H^{(l)}$ are local\cite{CEF}
integrals of motion 
\be
[H^{(m)},H^{(n)}]=0\ .
\ee
By local we mean that the densities of $H^{(m)}$ are local in
space. For fundamental spin models \cite{Korepinbook} they take the
form
\be
H^{(m)}=\sum_{j}H^{(m)}_{j,j+1,\ldots,j+m},
\ee
where $H^{(m)}_{j,j+1,\ldots,j+m}$ acts nontrivially only on sites
$j,j+1,\ldots,j+m$. 
The Lagrange multiplies $\lambda_l$ are fixed by the requirement that
the expectation values of the integrals of motion are time-independent
\be
\lim_{L\to\infty}\frac{\langle\Psi_0|H^{(l)}|\Psi_0\rangle}{L}
=\lim_{L\to\infty}\frac{{\rm Tr}\left[\rho_{\rm GGE}
H^{(l)}\right]}{L}.
\label{constraints}
\ee
Here $L$ is the size of the system under consideration. In practice it
is often useful to work with a truncated GGE \cite{FE:2013}, where only the
$y$ ``most local'' conservation laws are retained
\be
\rho^{(y)}_{\rm tGGE}=\frac{1}{Z^{(y)}_{\rm
    tGGE}}\exp\left(-\sum_{l=1}^y\lambda_l^{(y)}
H^{(l)}\right).
\label{tGGE}
\ee
Here the $\lambda^{(y)}_l$ are fixed by
\be
\lim_{L\to\infty}\frac{\langle\Psi_0|H^{(l)}|\Psi_0\rangle}{L}
=\lim_{L\to\infty}\frac{{\rm Tr}\left[\rho_{\rm tGGE}^{(y)}
H^{(l)}\right]}{L}\ ,\quad l=1,\ldots,y.
\ee
The full GGE is then recovered in the limit $y\to\infty$, after the
thermodynamic limit has been taken first. Assuming that a given
integrable system indeed approaches a stationary state late after a
quantum quench, which is described by a generalized Gibbs ensemble,
important questions are how to construct the GGE in practice, and how
to then determine expectation values of local operators. It is these
questions we aim to address for the particular case of the spin-1/2
Heisenberg XXZ chain. A priori there are four steps:
\begin{enumerate}
\item{} Determine the local conservation laws.
\item{} Calculate their expectation values in the initial state after
  the quench.
\item{} Construct the GGE density matrix in such a way
  that equations \fr{constraints} are fulfilled.
\item{} Determine the expectation values of local operators in this
  ensemble. 
\end{enumerate}
In the following we address these in turn. As local conservation laws
we consider the \emph{minimal set} obtained from the logarithmic derivative
of the transfer matrix at the ``shift-point'' \cite{Korepinbook}. It
has been recently found that the XXZ Hamiltonian in general has local
conservation laws that are not obtained in this way
\cite{nonlocal}. In principle they could be accommodated in our
construction as well. However, in order to keep things simple, we
restrict our analysis to the antiferromagnetically ordered regime of the
Heisenberg chain, where, as far as we know, the minimal set of local
conservation laws is complete. With regards to step 2, we focus on a
class of simple quenches, for which the initial states are unentangled.
We show how to treat these cases analytically. Our method generalizes
to weakly entangled initial states of matrix product form, but the
analysis becomes much more complicated\cite{FCCE}. The GGE density matrix is 
constructed by the quantum transfer matrix method \cite{QTM}. The most
difficult issue here is what values the Lagrange multipliers
$\lambda_j$ take. We argue that it is possible to completely specify
the quantum transfer matrix, without having to explicitly calculate the
$\lambda_j$. Finally, GGE expectation values of local operators can be
calculated by borrowing the results of the Wuppertal group for finite
temperature correlators \cite{wuppertal,wuppertal1}.

\section{Local integrals of motion}

We consider the XXZ Hamiltonian 
\be
H^{(1)}=\frac{1}{4}\sum_{\ell=1}^L \sigma_\ell^x\sigma_{\ell+1}^x+\sigma_{\ell}^y\sigma_{\ell+1}^y+\Delta(\sigma_\ell^z\sigma_{\ell+1}^z-1)\, ,
\label{HXXZ}
\ee
where $L$ is even, $\sigma^\alpha_j$ are Pauli matrices
$(\sigma_{L+1}^\alpha\equiv \sigma_1^\alpha)$ and we parametrize the 
anisotropy as 
\be
\Delta=\cos\gamma.
\ee
It is well known that
\fr{HXXZ} is solvable by the algebraic Bethe Ansatz method
\cite{Korepinbook}. Local conservation laws $H^{(k)}$ can then be
obtained from the logarithmic derivative of the transfer matrix
$\tau(i+\lambda)$ 
\be\label{eq:Hk}
H^{(k)}=i\Bigl(\frac{\sin \gamma}{\gamma}\frac{\partial}{\partial \lambda}\Bigr)^k\log \tau(i+\lambda)\Bigr|_{\lambda=0}\, .
\ee
By definition the conservation laws commute with one another
\be
[H^{(k)},H^{(n)}]=0.
\ee
The transfer matrix is constructed by Algebraic Bethe Ansatz and takes
the form
\bea
\label{eq:Lj}
\tau(i+\lambda)&=&{\rm Tr} \left[\L_L(\lambda)\L_{L-1}(\lambda)\ldots
\L_1(\lambda)\right] ,\nn
\L_j(\lambda)&=&\frac{1+\tau^z\sigma_j^z}{2}+\frac{\sinh(\frac{\gamma\lambda}{2})}{\sinh(i\gamma+\frac{\gamma\lambda}{2})}\frac{1-\tau^z\sigma_j^z}{2}+\frac{\sinh(i\gamma)}{\sinh(i\gamma+\frac{\gamma\lambda}{2})}(\tau^+\sigma_j^-+\tau^-\sigma_j^+),
\eea
where $\tau^{x,y,z}$ are Pauli matrices acting on the auxiliary space,
and the trace is taken over the latter. In the following we
denote indices in the auxiliary and quatum spaces by Roman ($a,b$) and
Greek ($\alpha,\beta$) letters respectively.

\section{Expectation values of local integrals of motion in the
  initial state}
Given an intial state $|\Psi_0\rangle$, we aim to determine the
expectation values
\be
\braket{\Psi_0|H^{(k)}|\Psi_0}\, .
\ee
It is convenient to work with the generating function
\be\label{eq:Omega}
\Omega_{\Psi_0}(\lambda)=\frac{1}{L}\braket{\Psi_0|\tau^\prime(i+\lambda)\tau^{-1}(i+\lambda)|\Psi_0}
=-i\sum_{k=1}
\left(\frac{\gamma}{\sin\gamma}\right)^k\frac{\lambda^{k-1}}{(k-1)!}
\frac{\langle\Psi_0|H^{(k)}|\Psi_0\rangle}{L},
\ee
where the right hand side follows from \fr{eq:Hk}. In order to
evaluate $\Omega_{\Psi_0}(\lambda)$ we use that, when viewed as a
power series in $\lambda$ for large $L$, we have formally
\be
\tau(i+\lambda)\sim\tau(i)\exp\left(-i\sum_{k=1}
\left(\frac{\gamma}{\sin\gamma}\right)^k
\frac{\lambda^k}{k!}H^{(k)}\right).
\ee
This suggests that for large $L$ we have
\be
\tau^{-1}(i+\lambda)=\left[\tau(i+\lambda)\right]^\dagger,
\ee
in the sense that the power-series expansions in $\lambda$ coincide.
These observations lead to the following (approximate) expression for
the inverse 
\bea
\label{eq:taum1}
\tau^{-1}(i+\lambda)&\sim&
{\rm Tr}\left[\M_L(\lambda)\M_{L-1}(\lambda)\ldots
\M_1(\lambda)\right] ,\\
\label{eq:Mj}
\M_j(\lambda)&=&\frac{1+\tau^z\sigma_j^z}{2}+\frac{\sinh(\frac{\gamma^\ast\lambda}{2})}{\sinh(-i\gamma^\ast+\frac{\gamma^\ast\lambda}{2})}\frac{1-\tau^z\sigma_j^z}{2}+\frac{\sinh(-i\gamma^\ast)}{\sinh(-i\gamma^\ast+\frac{\gamma^\ast\lambda}{2})}(\tau^+\sigma_j^++\tau^-\sigma_j^-)\, .
\eea
The generating function \fr{eq:Omega} can then be expressed as
\bea
\Omega_{\Psi_0}(\lambda)&=&\frac{1}{L}\frac{\partial}{\partial x}\Big|_{x=\lambda}
\langle\Psi_0|\tau(i+x)\tau^{-1}(i+\lambda)|\Psi_0\rangle\nn
&\sim&
\frac{1}{L}\frac{\partial}{\partial x}\Big|_{x=\lambda}{\rm Sp}
\langle\Psi_0|V_L(x,\lambda)\ldots
V_1(x,\lambda)|\Psi_0\rangle\ ,
\eea
where $V_n(x,\lambda)$ are $4\times 4$ matrices with entries 
$\left(V_n(x,\lambda)\right)^{ab}_{cd}$ that are
operators acting on the two-dimensional quantum space on site $n$
\be
\left(V_n(x,\lambda)\right)^{ab}_{cd}=(\L_n(x))^{ab}\left(\M_n(\lambda)\right)^{cd}.
\ee
In this notation $V_L(x,\lambda)\ldots V_1(x,\lambda)$ is a regular
product of $4\times 4$ matrices and $\rm{Sp}$ denotes the usual trace
for $4\times 4$ matrices. Let us now assume that $\ket{\Psi_0}$ is a
product state 
\be
\ket{\Psi_0}=\otimes_{j=1}^L\ket{ \Psi_0^{(j)}}.
\ee
Then $\Omega_{\Psi_0}$ can be written as
\be
\Omega_{\Psi_0}(\lambda)\sim
\frac{1}{L}\frac{\partial}{\partial x}\Big|_{x=\lambda}{\rm Sp}\left[
\prod_{j=1}^LU_j(x,\lambda)\right],
\label{omega_general}
\ee
where
\be
U_j(x,\lambda)=\langle\Psi_0^{(j)}|V_j(x,\lambda)|\Psi_0^{(j)}\rangle.
\label{Uj}
\ee
We now discuss how to implement the above programme for some explicit
examples. 

\subsection{Quench from $|x,\uparrow\rangle$}
\label{transverse}
Our first example is the product state
\be
|x,\uparrow\rangle=\otimes_{j=1}^L\frac{|\uparrow\rangle_j+|\downarrow\rangle_j}{\sqrt{2}}.
\label{productx}
\ee
This corresponds to all spins pointing in the x-direction. This initial
state corresponds to a quantum quench in the XXZ-chain with an applied
\emph{transverse} magnetic field
\be
H(h)=\frac{1}{4}\sum_{\ell=1}^L
\sigma_\ell^x\sigma_{\ell+1}^x+\sigma_{\ell}^y\sigma_{\ell+1}^y+\Delta(\sigma_\ell^z\sigma_{\ell+1}^z-1)-\frac{h}{2} \sum_{j=1}^L\sigma_j^x\, .
\ee
Preparing the system in the ground state of $H(\infty)$ gives the
initial state \fr{productx}, and the quench is to the integrable
zero-field Hamiltonian $H(0)$. Using translational invariance we have
\be
\Omega_{x,\uparrow}(\lambda)\sim
\frac{1}{L}\frac{\partial}{\partial x}\Big|_{x=\lambda}{\rm Sp}\left[
\left(U(x,\lambda)\right)^L\right].
\ee
Denoting the largest eigenvalue of $U(x,\lambda)$ by $\mu_{\rm
  max}(x,\lambda)$, this gives
\be
\Omega_{x,\uparrow}(\lambda)\sim\frac{1}{L}\frac{\partial}{\partial
  x}\Big|_{x=\lambda}\left(\mu_{\rm max}(x,\lambda)\right)^L=
\frac{\partial}{\partial x}\Big|_{x=\lambda}\mu_{\rm max}(x,\lambda).
\ee
In the last step we have used that $\mu_{\rm max}(\lambda,\lambda)=1$.
The matrix $U(x,\lambda)$ is readily calculated and its largest
eigenvalue, which for small $x,\lambda$ is very close to $1$, is
calculated using Mathematica. This results in 
\be
\Omega_{x,\uparrow}(\lambda)=
\frac{-i\gamma \sin(\gamma)}{2 + 2 \cosh(\gamma \lambda) + 4 \cos(\gamma)}\ .
\ee
Matching the power-series expansion around $\lambda=0$ with
\fr{eq:Omega} gives
\be
\lim_{N\to\infty}\frac{\braket{x,\uparrow|H^{(k)}|x,\uparrow}}{N}
=\frac{1-\Delta}{4}\frac{\partial^{k-1}}{\partial x^{k-1}}
\Bigr|_{x=0}\frac{ 1+\Delta}{\cosh^2(\sqrt{1-\Delta^2}x/2)+\Delta}.
\ee
The results for $H^{(1)}$, $H^{(2)}$ and $H^{(3)}$ agree with
Ref.~[\onlinecite{takahashi}].
\subsection{Quench from the N\'eel state}\label{Neel}
We next consider the N\'eel state
\be
|\text{\rm N\'eel}\rangle=\otimes_{j=1}^{L/2}
\Big[|\uparrow\rangle_{2j-1}\otimes|\downarrow\rangle_{2j}\Big].
\ee
This would be the initial state for a quantum quench from the ground
state of an initial XXZ Hamiltonian with $\Delta=+\infty$ to a final
Hamiltonian with finite $\Delta$. Using translational invariance by
two sites, our expression \fr{omega_general} for the generating
function becomes 
\be
\Omega_{\text{N\'eel}}(\lambda)\sim
\frac{1}{L}\frac{\partial}{\partial x}\Big|_{x=\lambda}{\rm Sp}\left[
\left(U_1(x,\lambda)U_2(x,\lambda)\right)^{L/2}\right],
\label{omega_Neel}
\ee
where
\bea
U_1(x,\lambda)&=&{}_1\langle\uparrow|V_1(x,\lambda)|\uparrow\rangle_1\ ,\nn
U_2(x,\lambda)&=&{}_2\langle\downarrow|V_2(x,\lambda)|\downarrow\rangle_2.
\eea
These two $4\times 4$ matrices and the largest eigenvalue of
$U_1(x,\lambda)U_2(x,\lambda)$ (for small $x,\lambda$) are readily
calculated using Mathematica

\be\label{eq:NS}
\Omega_{\text{N\'eel}}(\lambda)=\frac{i\gamma}{2}\frac{\sin(2\gamma)}
{2\cosh(\gamma\lambda)-1-\cos(2\gamma)}\, .
\ee
Matching the power series expansion around $\lambda=0$ with
\eqref{eq:Hk} then gives
\be
\lim_{L\to\infty}\frac{\braket{\text{N\'eel}|H^{(k)}|\text{N\'eel}}}{L}
=-\frac{\Delta}{2}\frac{\partial^{k-1}}{\partial x^{k-1}}
\Bigr|_{x=0}\frac{ 1-\Delta^2}{\cosh(\sqrt{1-\Delta^2}x)-\Delta^2}.
\ee

\subsection{More general initial states}
In principle our method can accommodate more complicated initial
states of matrix-product form
\be
|\Psi_0\rangle=\tilde{\rm Tr}\left[\otimes_{j=1}^L A_j\right],
\ee
where $A_j$ is an $m\times m$ matrix with entries that are quantum
states on site $j$, and $\tilde{\rm Tr}$ is the trace over the
$m$-dimensional matrix space. Considering such states is however
beyond the scope of the present paper; the generalization
to matrix product states is one of the subjects discussed in a
forthcoming publication~\cite{FCCE}. 
\section{Generalized Gibbs Ensemble and Quantum Transfer Matrix}

Combining \fr{tGGE} with \fr{eq:Hk}, we can express the density matrix
of the truncated generalized Gibbs ensembles as
\be
\rho^{(y)}_{\rm tGGE}=\frac{1}{Z^{(y)}_{\rm tGGE}} e^{-i \sum_{j=1}^y \lambda_j (\frac{\sin \gamma}{\gamma}\frac{\mathrm d}{\mathrm d x})^j \log \tau(i+x)\bigr|_{x=0}}\, .
\ee 
This density matrix can be analyzed by the quantum transfer matrix
approach \cite{QTM}. Following the analysis of Kl\"umper and Sakai
for computing the thermal conductivity at finite temperature
in Ref.~[\onlinecite{KS:2002}], we introduce inhomogeneities in the transfer
matrix and define the ensemble 
\be\label{eq:rhou}
\rho_{\{u_{1;N},\dots,u_{N;N}\}}=\frac{(\tau^{-1}(i)\tau(i+2u_{1;N}/\gamma))\cdots (\tau^{-1}(i)\tau(i+2u_{N;N}/\gamma))}{Z_{\{u_{1;N},\dots,u_{N;N}\}}}\, .
\ee
As transfer matrices with different spectral parameters commute
$[\tau(\lambda_1),\tau(\lambda_2)]=0$, we have 

\be
\rho_{\{u_{1;N},\dots,u_{N;N}\}}=\frac{e^{\sum_{i=1}^N\log \tau(i+2u_{1;N}/\gamma)-\log \tau(i)}}{Z_{\{u_{1;N},\dots,u_{N;N}\}}}\, .
\ee
In order to achieve (for asymptotically large $L$)
\be
\lim_{N\rightarrow\infty}\rho_{\{u_{1;N},\dots,u_{N;N}\}}=\rho^{(y)}_{\rm tGGE},
\ee
we need to choose the inhomogeneities $u_{j;N}$ such that
\be\label{eq:id}
\lim_{N\rightarrow\infty}\sum_{j=1}^N\bigl[\log \tau(i+2u_{j;N}/\gamma)-\log
\tau(i)\bigr]\overset{!}{=}-i\sum_{j=1}^y\lambda_j\Bigl(\frac{\sin
  \gamma}{\gamma} \frac{\mathrm d}{\mathrm d x}\Bigr)^j\log
\tau(i+x)\bigr|_{x=0}\ .
\ee
A sufficient condition for Eq.~\eqref{eq:id} to hold is that the
spectral parameters $u_{j;N}$ satisfy  
\be\label{eq:1}
\lim_{N\rightarrow\infty}\sum_{j=1}^N\bigl[f(i+2u_{j;N}/\gamma)-f(i)\bigr]=-i\sum_{j=1}^y
\lambda_j \Bigl(\frac{\sin\gamma}{\gamma}\frac{\mathrm d}{\mathrm d
  x}\Bigr)^j f(i+x)\Bigr|_{x=0} 
\ee
for any function $f(y)$ analytic at $y=i$. We do not have to require
further properties if we can find a solution such that 
\be\label{eq:Ninf}
\lim_{N\rightarrow\infty}u_{j;N}=0\, .
\ee
By series expanding $f$ about zero, Eq.~\eqref{eq:1} can be rewritten as
\be\label{eq:2}
\lim_{N\rightarrow\infty}\sum_{j=1}^N (u_{j;N})^\ell=
\begin{cases}
-i (\frac{\sin\gamma}{2})^\ell \ell!\ \lambda_\ell
& \text{if\ } \ell\leq y\\
0 & \text{if\ } \ell > y.
\end{cases}
\ee
A solution of \eqref{eq:2} satisfying \eqref{eq:Ninf} is given by 
\be
u_{m y+j;N}=u_{j;N}^{(y)}=\sum_{n=1}^y w_n^{(y)}\frac{e^{2\pi i j n/y}}{N^{n/y}}\qquad j=1,\hdots, y\quad m=0,\hdots,\lfloor N/y\rfloor -1\, ,
\ee
provided that
\be
\sum_{n_1,\hdots,n_\ell=1\atop \sum_{i=1}^\ell n_i=y}^{y+1-\ell} w_{n_1}^{(y)}\cdots w_{n_\ell}^{(y)}=-i \Bigl(\frac{\sin\gamma}{2}\Bigr)^\ell \ell! \lambda_\ell\, .
\ee
Having represented the density matrix of our truncated GGE in terms of
an inhomogeneous transfer matrix, we wish to generalize the quantum transfer
matrix approach for finite temperature correlation functions
[\onlinecite{GKS:2004}] in order to determine expectation values of
local operators in the (truncated) GGE. One can show that
for asymptotically large $L$
\be
Z_{\rm tGGE}^{(y)}\rho_{\rm tGGE}^{(y)}=\lim_{N\rightarrow\infty} \tr{\bar
  1,\hdots,\overline{2N}}{T_1^{QTM}(0)\cdots T_L^{QTM}(0)}\, , 
\ee
where the monodromy matrix $T_n^{QTM}$ acts on the physical space at
site $n$, and on the auxiliary spaces $\bar 1,\dots \overline{2N}$.
It is given by (we omit the subscript $n$, which only labels the site) 
\be
T^{QTM}(x)= \begin{pmatrix}
A(\gamma x/2)&B(\gamma x/2)\\
C(\gamma x/2)&D(\gamma x/2)
\end{pmatrix}=
\prod_{i=0}^{N-1}\L_{\overline{2N-i}}(x)\M_{\overline{2N-1-i}}\Bigl(\frac{2u_{i;N}}{\gamma}-x\Bigr)\, ,
\ee
and the matrices $\L$ and $\M$ are defined in
Eqs~\eqref{eq:Lj}\eqref{eq:Mj}. In the antiferromagnetic regime
$\Delta>1$ it is customary to define $\Delta=\cosh\eta$, corresponding
to $\eta=i\gamma$, and change variables as follows: 
\be
x=\frac{2}{\gamma}\lambda\, .
\ee
$T^{QTM}(2\lambda/\gamma)$ acts as an upper triangular matrix on the vector
\be
\ket{0}=\begin{pmatrix}1\\0\end{pmatrix}\otimes \begin{pmatrix}0\\1\end{pmatrix}\otimes\begin{pmatrix}1\\0\end{pmatrix}\otimes\cdots\otimes\begin{pmatrix}0\\1\end{pmatrix}
\ee
and we have
\be
C(\lambda)\ket{0}=0\, ,\qquad A(\lambda)\ket{0}=a(\lambda)\ket{0}\, ,\qquad D(\lambda)\ket{0}=d(\lambda)\ket{0}\, ,
\ee
with
\bea
a(\lambda)&=&\Bigl[\prod_{j=1}^y\frac{\sinh(\lambda-u_{j;N}^{(y)})}{\sinh(\lambda-u_{j;N}^{(y)}-\eta)}\Bigr]^{N/y}\ ,\nn
d(\lambda)&=&\Bigl(\frac{\sinh\lambda}{\sinh(\lambda+\eta)}\Bigr)^{N}\, .
\eea
The Bethe ansatz equations for the eigenvalues of the quantum transfer
matrix take the form
\be
\frac{a(w_j)}{d(w_j)}=\prod_{k=1\atop k\neq j}^M\frac{\sinh(w_j-w_k+\eta)}{\sinh(w_j-w_k-\eta)}\, ,
\ee
for some integer $M$, which, for the largest eigenvalue, is equal to $N$.
We then introduce the auxiliary function
\be\label{eq:adef}
\text{\textgoth{a}}(\lambda)=\frac{d(\lambda)}{a(\lambda)}\prod_{k=1}^{N}\frac{\sinh(\lambda-v_k+\eta)}{\sinh(\lambda-v_k-\eta)}\, ,
\ee
where $v_1,\dots,v_N$ is the solution of the Bethe ansatz equations
corresponding to the largest eigenvalue of the quantum transfer matrix
$t^{QTM}(x)={\rm Tr}[ T^{QTM}(x)]$. For $\Delta>1$ the function
$1+\text{\textgoth{a}}(\lambda)$ has simple zeros at $\lambda=v_j$ (on
the imaginary axis) and $y$ poles of order $N/y$ at
$\lambda=u_{j;N}^{(y)}$ inside the rectangle $Q$ defined by 
$|\mathrm{Re}\lambda|<\eta/2$ and $|\mathrm{Im} \lambda|<\pi/2$.
Using these analytic properties we obtain the following integral
equation for $\gotha(\lambda)$
\be
\log\text{\textgoth{a}}(\lambda)=\lim_{N\rightarrow\infty}\frac{N}{y}\sum_{j=1}^y\log\Big[\frac{\sinh(\lambda+\eta-u_{j;N}^{(y)})\sinh(\lambda)}{\sinh(\lambda-u_{j;N}^{(y)})\sinh(\lambda+\eta)}\Bigr]-\oint_{\mathcal C}\frac{\mathrm d \omega}{2\pi i}\frac{\sinh(2\eta)\log(1+\text{\textgoth a}(\omega))}{\sinh(\lambda-\omega+\eta)\sinh(\lambda-\omega-\eta)}\, ,
\ee 
where $\mathcal C$ is a rectangular contour with edges parallel to the
real axis at $\pm i\pi/2$ and to the imaginary axis at $\pm\gamma$
where $0<\gamma<\frac{\eta}{2}$. We may use Eq.~\eqref{eq:1} to
replace the inhomogeneities by the Lagrange multipliers $\lambda_j$
specifying the truncated GGE
\be
\lim_{N\rightarrow\infty}\frac{N}{y}\sum_{j=1}^y\log\Big[\frac{\sinh(\lambda+\eta-u_{j;N}^{(y)})\sinh(\lambda)}{\sinh(\lambda-u_{j;N}^{(y)})\sinh(\lambda+\eta)}\Bigr]=F\Bigl[i\frac{\sinh\eta}{2}\frac{\mathrm d }{\mathrm d \lambda}\Bigr]\Bigl(\frac{\sinh\eta}{\sinh\lambda\sinh(\lambda+\eta)}\Bigr),
\ee
where
\be
F[x]=-\frac{\sinh\eta}{2}\sum_{j=0}^{y-1}\lambda^{(y)}_{j+1} x^j\, .
\ee
Finally we have 
\be\label{eq:a}
\log\text{\textgoth{a}}(\lambda)=F\Bigl[i\frac{\sinh\eta}{2}\frac{\mathrm d }{\mathrm d \lambda}\Bigr]\Bigl(\frac{\sinh\eta}{\sinh\lambda\sinh(\lambda+\eta)}\Bigr)-\oint_{\mathcal C}\frac{\mathrm d \omega}{2\pi i}\frac{\sinh(2\eta)\log(1+\text{\textgoth a}(\omega))}{\sinh(\lambda-\omega+\eta)\sinh(\lambda-\omega-\eta)}\, .
\ee 
We stress that this equation is valid only if the expectation value of $S^z$
is zero. For non-vanising $\langle\Psi_0| S^z|\Psi_0\rangle$ one needs
to add a constant to the driving term, which corresponds to the
Lagrange multiplier of the total spin in the $z$ direction (see
\emph{e.g.} Ref.~[\onlinecite{GKS:2004}])  
\be\label{eq:ah}
\log\text{\textgoth{a}}(\lambda)=h+F\Bigl[i\frac{\sinh\eta}{2}\frac{\mathrm d }{\mathrm d \lambda}\Bigr]\Bigl(\frac{\sinh\eta}{\sinh\lambda\sinh(\lambda+\eta)}\Bigr)-\oint_{\mathcal C}\frac{\mathrm d \omega}{2\pi i}\frac{\sinh(2\eta)\log(1+\text{\textgoth a}(\omega))}{\sinh(\lambda-\omega+\eta)\sinh(\lambda-\omega-\eta)}\, .
\ee 
For the sake of simplicity we will restrict our analysis to the $h=0$
case in the following and return to the $h\neq 0$ in section~\ref{s:finiteh}.

The inverse function
$\bar\gotha(\lambda)\equiv 1/\gotha(\lambda)$ fulfils the following
integral equation
\be\label{eq:abar}
\log\bar{\text{\textgoth{a}}}(\lambda)=F\Bigl[i\frac{\sinh\eta}{2}\frac{\mathrm d }{\mathrm d \lambda}\Bigr]\Bigl(\frac{\sinh\eta}{\sinh(\lambda)\sinh(\lambda-\eta)}\Bigr)+\oint_{\mathcal C}\frac{\mathrm d \omega}{2\pi i}\frac{\sinh(2\eta)\log(1+\bar{\text{\textgoth a}}(\omega))}{\sinh(\lambda-\omega+\eta)\sinh(\lambda-\omega-\eta)}\, .
\ee
Similarly to the thermal case, the Lagrange multipliers enter into the
calculation of correlation functions only through the auxiliary
functions $\gotha(\lambda)$, $\bar\gotha(\lambda)$. 

\subsection{Thermodynamic Properties}
Thermodynamic properties are completely determined by the largest
eigenvalue $\Lambda_0(0)$ of the quantum transfer matrix. The
logarithm of $\Lambda_0(0)$ is given by~\cite{GKS:2004} 
\be
\log\Lambda_0(0)\equiv\lim_{L\rightarrow\infty}\lim_{N\rightarrow\infty}\frac{\log Z_{N,L}}{L}=\oint_{\mathcal C}\frac{\mathrm d \omega}{2\pi i}\frac{\sinh\eta\log(1+\gotha(\omega))}{\sinh\omega\sinh(\omega+\eta)}=-\oint_{\mathcal C}\frac{\mathrm d \omega}{2\pi i}\frac{\sinh\eta\log(1+\bar\gotha(\omega))}{\sinh\omega\sinh(\omega-\eta)}\, .
\ee
The expectation values of the local conservation laws can then be
expressed in the form
\be
\frac{1}{L}\braket{H^{(j)}}=-\frac{\partial \log\Lambda_0(0)}{\partial
  \lambda_j^{(y)}}=-\oint_{\mathcal C}\frac{\mathrm d \omega}{2\pi
  i}\frac{\sinh\eta\ \partial_{\lambda_j^{(y)}}\gotha(\omega)}{\sinh\omega\sinh(\omega+\eta)(1+\gotha(\omega))}=\oint_{\mathcal
  C}\frac{\mathrm d \omega}{2\pi
  i}\frac{\sinh\eta(\partial_{\lambda_j^{(y)}}\log\bar\gotha(\omega))}{\sinh\omega\sinh(\omega-\eta)(1+\bar \gotha(\omega))}\,
. 
\label{Hj}
\ee
Using \eqref{eq:abar}, we obtain
\be\label{eq:derabar}
\partial_{\lambda_j^{(y)}}\log\bar{\text{\textgoth{a}}}(\lambda)=-\Bigl(i\frac{\sinh\eta}{2}\Bigr)^{j-1}\partial^{j-1}_\lambda\Bigl(\frac{\sinh^2\eta}{2\sinh(\lambda)\sinh(\lambda-\eta)}\Bigr)+\oint_{\mathcal C}\frac{\mathrm d \omega}{2\pi i}\frac{\sinh(2\eta)(\partial_{\lambda_j^{(y)}}\log\bar\gotha(\omega))}{\sinh(\lambda-\omega+\eta)\sinh(\lambda-\omega-\eta)(1+\gotha(\omega))}\, .
\ee
The solution to this integral equation is conveniently expressed in
terms of an auxiliary function $G$
\be
\partial_{\lambda_j^{(y)}}\log\bar{\text{\textgoth{a}}}(\lambda)=-i\Bigl(-i\frac{\sinh\eta}{2}\Bigr)^{j}\partial_\mu^{j-1}G(\lambda,\mu;0)\Bigr|_{\mu=0},
\ee
where 
\bea
\label{eq:G1}
G(\lambda,\mu;\alpha)&=&-\coth(\lambda-\mu)+e^{\alpha\eta}\coth(\lambda-\mu-\eta)+\oint_{\mathcal
  C}\frac{\mathrm d \omega}{2\pi
  i}\frac{G(\omega,\mu;\alpha)}{1+\text{\textgoth{a}}(\omega)}K(\lambda-\omega;\alpha)\ ,\nn
K(\lambda;\alpha)&=&e^{\alpha \eta}\coth(\lambda-\eta)-e^{-\alpha\eta}\coth(\lambda+\eta)\, .
\eea
Substituting this back into \fr{Hj}, we obtain
\be
\frac{1}{L}\braket{H_j}=-i\Bigl(\frac{-i \sinh\eta}{2}\Bigr)^{j}\partial_\mu^{j-1}\oint_{\mathcal C}\frac{\mathrm d \omega}{2\pi i}\frac{\sinh\eta\ G(\omega,\mu;0)}{\sinh\omega\sinh(\omega-\eta)(1+\gotha(\omega))}\Bigr|_{\mu=0}\, .
\label{Hj2}
\ee

For the case of the full generalized Gibbs ensemble, i.e. the limit
$y\to\infty$ of the truncated GGE considered above, we may lift the
relationship \fr{Hj2} to the level of the generating function
$\Omega_{\Psi_0}(\mu)$ defined in \fr{eq:Omega} 
\be
\partial_\mu^{j-1}\Bigl[\oint_{\mathcal C}\frac{\mathrm d \omega}{2\pi
    i}\frac{\sinh\eta\ G(\omega,\mu;0)}{\sinh\omega\sinh(\omega-\eta)(1+\gotha(\omega))}+\Bigl(\frac{2i}{\eta}\Bigr)^{j}\Omega_{\Psi_0}(\mu)\Bigr]
\Bigr|_{\mu=0}=0\ ,
\qquad j=1,2,\hdots\, .
\ee
This can be rewritten in a more compact form as
\be
\oint_{\mathcal C}\frac{\mathrm d \omega}{2\pi
  i}\frac{\sinh\eta\ G(\omega,\mu;0)}{\sinh\omega\sinh(\omega-\eta)(1+\gotha(\omega))}=-\frac{2i}{\eta}\
\Omega_{\Psi_0}\left(\frac{2 i \mu}{\eta}\right).
\label{Gomega}
\ee

In passing we note that the magnetization can be computed analogously
by taking the derivative with respect to the magnetic field of the
integral equation~\eqref{eq:ah}, which gives 
\be
\oint_{\mathcal C}\frac{\mathrm d \omega}{2\pi i}\frac{G(\omega,0;0)}{1+\gotha(\omega)}=-\frac{\braket{\sigma^z}+1}{2}\, .
\ee

To summarize the result of this section: in the thermodynamic limit
the GGE can be formulated in terms of a quantum transfer matrix, whose
largest eigenvalue is given in terms of the functions $\gotha$ and
$\bar{\gotha}$, which in turn are defined through the nonlinear
integral equations \fr{eq:a} and \fr{eq:abar} respectively. The
expectation values of the local integrals of motion can then be expressed in
terms of $\gotha$ and the auxiliary function $G(\lambda,\mu;0)$ defined
in \fr{eq:G1}. In order to proceed for the case of a truncated GGE, it
is necessary to solve our system of integral equations subject to the
constraints \fr{Hj2}. Carrying out such a computation entails
calculating the Lagrange multipliers $\lambda_j^{(y)}$. On the other
hand, we are ultimately interested 
in the full GGE itself. In this case it is possible to avoid having to
determine the Lagrange multipliers $\lambda_j^{(y)}$, as we will show next.

\section{Eliminating the Lagrange multipliers}\label{s:elLag}
In order to proceed, it is convenient to switch to the ``$\gothb
\bar\gothb$-formulation'' of the integral equations by defining
\bea
\label{eq:defb}
\gothb(x)&=&\gotha\Bigl(i x^++\frac{\eta}{2}\Bigr)\ ,
\quad \bar \gothb(x)=\bar \gotha\Bigl(i x^--\frac{\eta}{2}\Bigr)\ ,\nn
g_\mu^\pm(x)&=& \pm G\Bigl(i x^{\pm}\pm\frac{\eta}{2},i\mu;0\Bigr)\ ,
\quad x^{\pm}=x\pm i\epsilon\, .
\eea
The functions~\eqref{eq:defb} can be shown to obey the integral equations
\be\label{eq:system1}
\ba
{}
\log\gothb(x)&=
F\Bigl[\frac{\sin\eta}{2} \frac{\partial}{\partial x}\Bigr]
d(x)+[k\ast\log(1+\gothb)](x)-[k_-\ast\log(1+\bar\gothb)](x)\ ,\\
\log\bar\gothb(x)&=F\Bigl[\frac{\sin\eta}{2} \frac{\partial}{\partial
    x}\Bigr]
d(x)+[k\ast\log(1+\bar\gothb)](x)-[k_+\ast\log(1+\gothb)](x)\ ,\\
g_\mu^+(x)&=-d(x-\mu)+\Bigl[k\ast\frac{g_\mu^+}{1+\gothb^{-1}}\Bigr](x)-\Bigl[k_-\ast\frac{g_\mu^-}{1+\bar{\gothb}^{-1}}\Bigr](x)\ ,\\
g_\mu^-(x)&=-d(x-\mu)+\Bigl[k\ast\frac{g_\mu^-}{1+\bar\gothb^{-1}}\Bigr](x)-\Bigl[k_+\ast\frac{g_\mu^+}{1+\gothb^{-1}}\Bigr](x)
\ea
\ee
where $[f_1\ast f_2](x)=\int_{-\pi/2}^{\pi/2}\frac{\mathrm dy}{\pi}
f_1(x-y)f_2(y)$ and
\be
d(x)=\sum_{n=-\infty}^{\infty}\frac{e^{2 i n x}}{\cosh(\eta n)}\ ,
\quad k(x)=\sum_{n=-\infty}^{\infty}\frac{e^{2 i n x}}{e^{2\eta
    |n|}+1}\ ,
\quad k_\pm(x)=k(x^\mp \pm i\eta)\, .
\label{dk}
\ee
As shown in Refs~[\onlinecite{wuppertal}], $d(x)$ and $k(x)$ can be
expressed in terms of special functions. In terms of the new
variables, equation \fr{Gomega} is rewritten as
\be\label{eq:constraint}
-\int_{-\frac{\pi}{2}}^{\frac{\pi}{2}}\frac{\mathrm d x}{\pi}d(x)\Bigl(\frac{g^+_{\mu}(x)}{1+\gothb^{-1}(x)}+\frac{g^-_{\mu}(x)}{1+\bar\gothb^{-1}(x)}\Bigr)=4 k(\mu)+\frac{4 i}{\eta}\Omega_{\Psi_0}(-2\mu/\eta)\, .
\ee
The Lagrange multipliers $\lambda_j$ are just parameters of the
integral equations, which are fixed by Eq.~\eqref{eq:constraint}. 
Quite surprisingly, we can remove the explicit dependence on
$\lambda_j$ by taking the difference between the first two equations
of system~\eqref{eq:system1}, i.e. considering the system of integral
equations 
\bea
\label{eq:system2}
\log\gothb(x)-\log\bar\gothb(x)&=&[(k_++k)\ast\log(1+\gothb)](x)-[(k_-+k)\ast\log(1+\bar\gothb)](x)\ ,\nn
g_\mu^+(x)&=&-d(x-\mu)+\Bigl[k\ast\frac{g_\mu^+}{1+\gothb^{-1}}\Bigr](x)-\Bigl[k_-\ast\frac{g_\mu^-}{1+\bar{\gothb}^{-1}}\Bigr](x)\ ,\nn
g_\mu^-(x)&=&-d(x-\mu)+\Bigl[k\ast\frac{g_\mu^-}{1+\bar\gothb^{-1}}\Bigr](x)-\Bigl[k_+\ast\frac{g_\mu^+}{1+\gothb^{-1}}\Bigr](x)\ ,\nn
4 k(\mu)+\frac{4 i}{\eta}\Omega_{\Psi_0}(-2\mu/\eta)&=&
-\int_{-\frac{\pi}{2}}^{\frac{\pi}{2}}\frac{\mathrm d
  x}{\pi}d(x)\Bigl(\frac{g^+_{\mu}(x)}{1+\gothb^{-1}(x)}+\frac{g^-_{\mu}(x)}{1+\bar\gothb^{-1}(x)}\Bigr)\ .
\eea
In general \fr{eq:system2} have to be solved numerically by iteration.
In addition, from the the same equations it follows that
\be\label{eq:Lmult}
F[i n\sinh\eta]=\frac{e^{-\eta n}[\log\gothb]_n+e^{\eta n}[\log\bar\gothb]_n}{2}\, ,
\ee
where
\be
[f]_n=\int_{-\frac{\pi}{2}}^{\frac{\pi}{2}}\frac{\mathrm d x}{\pi}e^{-2i n x}f(x)\, .
\ee
Let us briefly comment on how to solve the system \fr{eq:system2}. The
second and third equations are linear in $g_\mu^{\pm}$ and can be
straightforwardly inverted in order to express $g_\mu^{\pm}$ as
functions of $\gothb$ and $\bar\gothb$. The first equation of
\fr{eq:system2} is nonlinear, but it is reasonable to expect that
it can be used to express $\bar \gothb$ in terms of $\gothb$. The last
equation, which is more conveniently analyzed in Fourier space, can
finally be inverted to obtain the remaining unknown $\gothb$. 
We have analyzed a variety of initial states\cite{FCCE} and find that
such an iteration scheme appears to always converge to a unique
solution. 

By setting $n=0$ in \eqref{eq:Lmult}, one obtains the inverse
temperature (the Lagrange multiplier of the Hamiltonian). 
The other
Lagrange multipliers are more difficult to compute, as it is not
always easy to identify the Taylor coefficients of $F[x]$
from the values of the Fourier coefficients on the right hand side
(the case considered in Section~\ref{s:NS} is an example). 
This suggests that solving the system~\eqref{eq:system2} is less
demanding than working out \eqref{eq:system1} by recursively computing
the Lagrange multipliers.

\section{Perturbative expansion}\label{s:pert}
We now consider examples, in which \fr{eq:system2} can be solved
through an expansion that can be carried out
analytically. Importantly, this shows that it is indeed possible 
to completely specify the quantum transfer matrix describing the GGE
without having to calculate the Lagrange multipliers $\lambda_j$.

If we consider the expectation value on the ground state of the model,
it is known that  
\be
\Omega_{GS}(-2\mu/\eta)=i \eta k(\mu)\, ,
\label{OGS}
\ee
where $k(\mu)$ is given in \fr{dk}.
Eqn \fr{OGS} can be inferred directly from \eqref{eq:system2},
\eqref{eq:Lmult} by considering the case
$\gothb(x)=\epsilon(1+\mathcal O(\epsilon^2))$ and
$\bar\gothb(x)=\epsilon(1+\mathcal O(\epsilon^2))$ in the limit 
$\epsilon\to 0$. This is equivalent to the zero temperature limit
in the Gibbs ensemble, which in the $\Delta>1$ case clearly tends to
the thermodynamic ground state of the model (see also
Ref.~[\onlinecite{GKS:2004}]).

This observation opens the door for carrying out a perturbative
expansion of the system~\eqref{eq:system2} in the limit of a ``small''
quench. We define functions $\rho(x)$ and $\zeta(x)$ by
\bea
\frac{1}{1+\gothb^{-1}(x)}&=&e^{\zeta(x)/2}\rho(x)\ ,\nn
\frac{1}{1+\bar\gothb^{-1}(x)}&=&e^{-\zeta(x)/2}\rho(x)\, .
\label{rhozeta}
\eea
In terms of these new variables, \eqref{eq:system2} are rewritten as
\be\label{eq:system3}
\begin{gathered}
{}
\zeta(x)=-[(k+k_+)\ast\log(1-e^{\zeta/2}\rho)](x)+[(k+k_-)\ast\log
  (1-e^{-\zeta/2}\rho)](x)+\log\Bigl(\frac{1-e^{\zeta(x)/2}\rho(x)}{1-e^{-\zeta(x)/2}\rho(x)}\Bigr)\ ,\\
g_\mu^+(x)=-d(x-\mu)+\Bigl[k\ast(\rho e^{\zeta/2}
  g_\mu^+)\Bigr](x)-\Bigl[k_-\ast(\rho e^{-\zeta/2}
  g_\mu^-)\Bigr](x)\ ,\\
g_\mu^-(x)=-d(x-\mu)+\Bigl[k\ast(\rho
  e^{-\zeta/2}g_\mu^-)\Bigr](x)-\Bigl[k_+\ast(\rho
  e^{\zeta/2}g_\mu^+)\Bigr](x)\ ,\\
\int_{-\frac{\pi}{2}}^{\frac{\pi}{2}}\frac{\mathrm d x}{4\pi}d(x)\rho(x)\Bigl[ e^{\zeta(x)/2}g_{\mu}^+(x)+e^{-\zeta(x)/2}g_{\mu}^-(x)\Bigr]=-\frac{i}{\eta}\Omega_{\Psi_0}(-2\mu/\eta)-k(\mu)\, .
\end{gathered}
\ee
The definition of a ``small'' quench is one for which $|\rho(x)|\ll 1$.
In this case we may solve \fr{eq:system3} by iteration. At lowest
order we have 
\be
\rho(x)\sim \rho^{(1)}(x),\quad \zeta(x)\sim 0\ ,\quad
g_\mu^\pm(x)\sim -d(x-\mu).
\ee
At this order, the last equation of \eqref{eq:system2} reads
\be
[d\ast (\rho^{(1)}\, d)](\mu)=2k(\mu)+\frac{2 i}{\eta}\Omega_{\Psi_0}(-2\mu/\eta)	\, .
\ee
This can be solved by Fourier techniques. The n'th Fourier coefficient is
\be
\frac{[\rho^{(1)}\, d]_n}{\cosh(\eta n)}=\frac{2}{1+e^{2\eta|n|}}+\frac{2i}{\eta}\int_{-\frac{\pi}{2}}^{\frac{\pi}{2}}\frac{\mathrm d \mu}{\pi}e^{-2i n\mu}\Omega_{\Psi_0}(-2\mu/\eta)\, ,
\ee
and going back to real space we obtain
\be\label{eq:rho1}
\rho^{(1)}(x)=\frac{\frac{i}{\eta}\Omega_{\Psi_0}(-\frac{2x}{\eta}+i)+\frac{i}{\eta}\Omega_{\Psi_0}(-\frac{2x}{\eta}-i)+K_\eta(x)}{d(x)}\, ,
\ee
where
\be\label{eq:Keta}
K_\eta(x)=\frac{\sinh\eta}{\cosh\eta-\cos(2x)}\, .
\ee
In the next order of iteration one replaces $\rho(x)$ with
$\rho^{(1)}(x)$ in the equations defining $\zeta(x)$ and
$g_\mu^\pm(x)$, and solving the resulting system gives the improved
result $\rho^{(2)}(x)$
\be\label{eq:rho2}
\rho^{(2)}(x)=\rho^{(1)}(x)\left\{1-\frac{1}{d(x)}\Bigl[\Bigl(k-\frac{1}{4}K_{2\eta}\Bigr)\ast(\rho^{(1)} d )\Bigr](x)\right\}\, .
\ee
The parameter
\be\label{eq:kappa}
\kappa\equiv \max_{x}\left|\frac{1}{d(x)}\Bigl[\Bigl(k-\frac{1}{4}K_{2\eta}\Bigr)\ast(\rho^{(1)} d )\Bigr](x)\right|
\ee
gives an estimate for the accuracy of the first order approximation.
We will show in Section~\ref{s:cfun} that for a small quench the entire
dependence on the initial state is encoded in
$\rho^{(1)}(x)$. Short-range spin-spin correlators are then expressed
in terms of integrals involving $\rho^{(1)}(x)$ and other known
functions, which depend \emph{only} on the anisotropy parameter $\Delta$.

\subsection{Example: quench from the N\'eel state ($\Delta_0=+\infty$)
to finite, large $\Delta$}\label{s:NS}

In this section we consider a quench from the N\'eel state, for which 
$\Omega_{\text{N\'eel}}(\mu)$ is given in \eqref{eq:NS}. The lowest
order result for $\rho^{(1)}(x)$ \eqref{eq:rho1} is given by
\be\label{eq:rho1NS}
\rho^{(1)}(x)=\frac{\sin^2(2x)\tanh^2(\eta)\sinh(\eta)}{[\cosh(\eta)-\cos(2x)][(\cosh(\eta)-\cos(2x))^2+\tanh^2(\eta)\sin^2(2x)]d(x)}\xrightarrow{1\ll\eta} \frac{\sin^2(2x)}{\cosh^2(\eta)}\, .
\ee
This is proportional to $1/\Delta^2$ and for sufficiently large
$\Delta$ indeed small.

\section{Correlation functions}\label{s:cfun}

In the last years there has been tremendous progress in the calculation of equal time correlation functions in the spin-1/2 XXZ
chain~\cite{wuppertal,wuppertal1,GKS:2004,Smirnov}. These results can be applied also to the generalized Gibbs ensemble of interest here. In particular, we may use 
the explicit expressions for short-distance correlators given in Ref.~[\onlinecite{wuppertal1}],
which involve three functions
$\varphi(\mu)$, $\omega(\mu_1,\mu_2)$ and
$\omega^\prime(\mu_1,\mu_2)$. 
Examples are
\be
\langle\sigma^z_1\sigma^z_2\rangle={\rm
  cth}(\eta)\omega+\frac{\omega'_x}{\eta}\ ,\quad
\langle\sigma^x_1\sigma^x_2\rangle=-\frac{\omega}{2\sinh(\eta)}-
\frac{\cosh(\eta)\omega'_x}{2\eta}\ ,
\label{zzxx}
\ee
where $\omega=\omega(0,0)$, $\omega'_x=\partial_x\omega'(x,y)|_{x,y=0}$.
The corresponding expressions for the GGE
are identical, but the functions
$\varphi(\mu)$, $\omega(\mu_1,\mu_2)$, $\omega^\prime(\mu_1,\mu_2)$
are different. Following the finite temperature case, we define
functions
\be
g_\mu^{\prime \pm}(x)= \pm \partial_\alpha G\Bigl(i x^{\pm}\pm\frac{\eta}{2},i\mu;\alpha\Bigr)\Bigr|_{\alpha=0}\, .
\ee
It follows from Eq.~\eqref{eq:G1} that $g_\mu^{\prime \pm}(x)$ satisfy
the integral equations 
\be
\ba
g_\mu^{\prime +}(x)&=-\eta c_+(x-\mu)+\eta\Bigl[\ell\ast
  \frac{g^+_\mu}{1+\gothb^{-1}}\Bigr](x)-\eta\Bigl[\ell_-\ast
  \frac{g^-_\mu}{1+\bar\gothb^{-1}}\Bigr](x)+\Bigl[\kappa\ast\frac{g^{\prime+}_\mu}{1+\gothb^{-1}}\Bigr](x)-\Bigl[\kappa_-\ast\frac{g^{\prime-}_\mu}{1+\bar\gothb^{-1}}\Bigr](x)\ ,\\
g_\mu^{\prime -}(x)&=-\eta c_-(x-\mu)+\eta\Bigl[\ell\ast \frac{g^-_\mu}{1+\bar\gothb^{-1}}\Bigr](x)-\eta\Bigl[\ell_+\ast \frac{g^+_\mu}{1+\gothb^{-1}}\Bigr](x)+\Bigl[\kappa\ast\frac{g^{\prime-}_\mu}{1+\bar \gothb^{-1}}\Bigr](x)-\Bigl[\kappa_+\ast\frac{g^{\prime+}_\mu}{1+\gothb^{-1}}\Bigr](x)\, ,
\ea
\ee
where
\be
\ell(x)=\sum_{n=-\infty}^{\infty}\frac{\mathrm{sgn}(n)e^{2i n
    x}}{4\cosh^2(\eta n)}\, ,\quad \ell_\pm(x)=\ell(x^{\mp}\pm
i\eta)\ ,\quad
c_{\pm}(x)=\pm\sum_{n=-\infty}^{\infty}\frac{e^{\pm \eta n+2i n x}}{2\cosh^2(\eta n)}\, .
\ee
The three functions entering the expressions for short-range
correlators are then given by
\be\label{eq:Smirnov}
\begin{gathered}
\varphi(\mu)=\int_{-\frac{\pi}{2}}^{\frac{\pi}{2}}\frac{\mathrm d x}{2\pi}\Bigl(\frac{g_{-i \mu}^{-}(x)}{1+\bar\gothb^{-1}(x)}-\frac{g_{-i \mu}^{+}(x)}{1+\gothb^{-1}(x)}\Bigr)\\
\omega(\mu_1,\mu_2)=\omega_{(0)}(\mu_1,\mu_2)-\Bigl[d\ast\Bigl(\frac{g_{-i\mu_1}^+(x)}{1+\gothb^{-1}(x)}+\frac{g_{-i\mu_1}^-(x)}{1+\bar \gothb^{-1}(x)}\Bigr)\Bigr](-i\mu_2)\\
\omega^\prime(\mu_1,\mu_2)=\omega^\prime_{(0)}(\mu_1,\mu_2)-\Bigl[d\ast\Bigl(\frac{g_{-i\mu_1}^{\prime +}}{1+\gothb^{-1}}+\frac{g_{-i\mu_1}^{\prime -}}{1+\bar \gothb^{-1}}\Bigr)\Bigr](-i\mu_2)-\eta \Bigl[c_-\ast\frac{g_{-i\mu_1}^+}{1+\gothb^{-1}}\Bigr](-i\mu_2)-\eta\Bigl[c_+\ast\frac{g^-_{-i\mu_1}}{1+\bar\gothb^{-1}}\Bigr](-i\mu_2)\, ,
\end{gathered}
\ee
where
\be
\ba
\omega_{(0)}(\mu_1,\mu_2)&=-4 k(i\mu_1-i\mu_2)+K_{2\eta}(i\mu_1-i\mu_2)\\
\omega^\prime_{(0)}(\mu_1,\mu_2)&=-4\eta\ell(i\mu_1-i\mu_2)+\eta K_{2(\mu_1-\mu_2)}(i\eta)\, .
\ea
\ee

\subsection{Perturbative results for ``small'' quenches}\label{ss:NScorr}
For small quenches as defined in section \ref{s:pert} we may use the
iterative solution of the integral equations for $\gothb$ to obtain
approximate expressions for the functions $\varphi$, $\omega$ and $\omega'$.
Using the parametrization \fr{rhozeta} and the lowest order
solution \eqref{eq:rho1}, we find the following expressions for
the first order approximations to the functions~\eqref{eq:Smirnov}
\bea\label{eq:corrfirst}
\varphi_{(1)}(\mu)&=&0\ ,\nn
\omega_{(1)}(\mu_1,\mu_2)&=&\omega_{(0)}(\mu_1,\mu_2)+2\int_{-\frac{\pi}{2}}^{\frac{\pi}{2}}\frac{\mathrm
  d x}{\pi} d(x+i\mu_1)d(x+i\mu_2)\rho^{(1)}(x)\ ,\nn
\omega^\prime_{(1)}(\mu_1,\mu_2)&=&\omega^\prime_{(0)}(\mu_1,\mu_2)+\eta\int_{-\frac{\pi}{2}}^{\frac{\pi}{2}}\frac{\mathrm
  d x}{\pi} d(x+i\mu_2)\rho^{(1)}(x)\bar
c(x+i\mu_1)-\eta\int_{-\frac{\pi}{2}}^{\frac{\pi}{2}}\frac{\mathrm d
  x}{\pi} \bar c(x+i\mu_2)\rho^{(1)}(x)d(x+i\mu_1)\ ,
\label{3fnsapprox}
\eea
where we have defined $\bar c(x)=c_+(x)+c_-(x)$. We can determine
the functions entering the expressions for short-distance correlation
functions such as \fr{zzxx} by numerical integration of the
appropriate partial derivatives of \fr{3fnsapprox} at $(0,0)$, once
the function $\rho^{(1)}$ is known. 

Results for a quench from the
N\'eel state, where $\rho^{(1)}(x)$ is given by~\eqref{eq:rho1NS}, are
presented in Table~\ref{t:1}. We stress that these are not obtained through
a numerical solution of the nonlinear integral equations; we have
merely evaluated the analytic
expressions~\eqref{eq:rho1NS}\eqref{eq:corrfirst}.  
To the best of our knowledge, these are the first closed-form
expressions for correlators in the stationary state after a nontrivial
interacting quench.

\begin{table}
\begin{center}
\begin{tabular}{cc}
\begin{tabular}[b]{c||c|c|c}
$\Delta=2\atop(\kappa\sim0.1)$&$\ell$&$\braket{\sigma_1^z\sigma_{1+\ell}^z}$&$\braket{\sigma_1^x\sigma_{1+\ell}^x}$\\
\hline
\hline
&1&-0.661371&-0.338629\\
&2&0.376573&0.056895\\
&3&-0.279034&-0.009872
\end{tabular}&
\begin{tabular}[b]{c||c|c|c}
$\Delta=3\atop(\kappa\sim0.03)$&$\ell$&$\braket{\sigma_1^z\sigma_{1+\ell}^z}$&$\braket{\sigma_1^x\sigma_{1+\ell}^x}$\\
\hline
\hline
&1&-0.815293&-0.27706\\
&2&0.643582&0.039557\\
&3&-0.57034&-0.005788
\end{tabular}\\
\begin{tabular}{c||c|c|c}
$\Delta=4\atop(\kappa\sim0.01)$&$\ell$&$\braket{\sigma_1^z\sigma_{1+\ell}^z}$&$\braket{\sigma_1^x\sigma_{1+\ell}^x}$\\
\hline
\hline
&1&-0.887611&-0.224779\\
&2&0.779648&0.025859\\
&3&-0.730205&-0.003019
\end{tabular}&
\begin{tabular}{c||c|c|c}
$\Delta=5\atop(\kappa\sim0.008)$&$\ell$&$\braket{\sigma_1^z\sigma_{1+\ell}^z}$&$\braket{\sigma_1^x\sigma_{1+\ell}^x}$\\
\hline
\hline
&1&-0.925316&-0.186711\\
&2&0.852509&0.0177252\\
&3&-0.81808&-0.001699
\end{tabular}
\end{tabular}
\caption{The short-range correlators for various values of $\Delta$
after a quench from the N\'eel state ($\Delta_0=+\infty$) at lowest
order of the perturbative expansion; $\kappa$ \eqref{eq:kappa} is an
estimate of the relative error for $\rho^{(1)}(x)$.}\label{t:1} 
\end{center}
\end{table}

\section{Initial states with nonzero longitudinal magnetization}
\label{s:finiteh}
If the initial state has a nonzero magnetization
\be
m^z=\frac{1}{2L}\sum_{\ell}\braket{\sigma_\ell^z}\neq 0
\ee
the integral equations~\eqref{eq:system2} must be modified as follows
\be
\ba
{}&\log\gothb(x)-\log\bar\gothb(x)+h=[(k_++k)\ast\log(1+\gothb)](x)-[(k_-+k)\ast\log(1+\bar\gothb)](x)\ ,\\
&g_\mu^+(x)=-d(x-\mu)+\Bigl[k\ast\frac{g_\mu^+}{1+\gothb^{-1}}\Bigr](x)-\Bigl[k_-\ast\frac{g_\mu^-}{1+\bar{\gothb}^{-1}}\Bigr](x)\ ,\\
&g_\mu^-(x)=-d(x-\mu)+\Bigl[k\ast\frac{g_\mu^-}{1+\bar\gothb^{-1}}\Bigr](x)-\Bigl[k_+\ast\frac{g_\mu^+}{1+\gothb^{-1}}\Bigr](x)\ ,\\
&-\int_{-\frac{\pi}{2}}^{\frac{\pi}{2}}\frac{\mathrm d
  x}{\pi}d(x)\Bigl(\frac{g^+_{\mu}(x)}{1+\gothb^{-1}(x)}+\frac{g^-_{\mu}(x)}{1+\bar\gothb^{-1}(x)}\Bigr)=4
k(\mu)+\frac{4 i}{\eta}\Omega_{\Psi_0}(-2\mu/\eta)\ ,\\
&\int_{-\frac{\pi}{2}}^{\frac{\pi}{2}}\frac{\mathrm d x}{\pi}\Bigl(\frac{g_{0}^{+}(x)}{1+\gothb^{-1}(x)}-\frac{g_{0}^{-}(x)}{1+\bar\gothb^{-1}(x)}\Bigr)=4m^z\, ,
\ea
\ee
where $h$ is the Lagrange multiplier of the total spin in the
longitudinal direction. 

For a small quench, for which the magnetization is close to zero, the
right hand side of the last two equations is still small, so we can
still use the approach developed in Section~\ref{s:pert}. The main
difference is that at lowest order the function $\zeta(x)$ is now a
nonzero constant. The modification is in fact trivial, and we easily
obtain 
\be
\sinh(\zeta/2)\sim
-\frac{2m^z}{\int_{-\frac{\pi}{2}}^{\frac{\pi}{2}}\frac{\mathrm d
    x}{\pi}d(x)\rho^{(1)}_0(x)}\ ,\quad
\rho^{(1)}(x)\sim\frac{\rho^{(1)}_0(x)}{\cosh(\zeta/2)},
\ee
where $\rho^{(1)}_0(x)$ is the function \eqref{eq:rho1}. 
As a matter of fact, at the lowest order $\omega_{(1)}(\mu_1,\mu_2)$ and $\omega_{(1)}^\prime(\mu_1,\mu_2)$ have the same form in terms of the generating function $\Omega_{\Psi_0}$ as for $m^z=0$. The only difference is in $\varphi(\mu)$, which is now different from zero
\be
\varphi_{(1)}(\mu)=\tanh(\zeta/2)[d\ast\rho^{(1)}_0](-i\mu)\, .
\ee

\section{Summary and Conclusions}

We have considered the problem of the late time behaviour of
short-distance spin-spin correlators in the spin-1/2 Heisenberg XXZ
chain after certain quantum quenches. We focussed on the
antiferromagnetically ordered regime of the XXZ chain and assumed that
at infinite times a generalized Gibbs ensemble is reached. Following
Ref.~[\onlinecite{KS:2002}] we constructed quantum transfer matrix
description of this ensemble, and derived a set of nonlinear integral
equations that describe the largest eigenvalue of the quantum transfer
matrix. We emphasize that there is an important difference between our
description of the stationary state after a quench and thermal
ensembles for generalized integrable Hamiltonians of the kind considered in
Ref.~[\onlinecite{KS:2002}]: in our case the ``input data'' are not
given by the Lagrange multipliers (inverse temperature, chemical
potential, etc.), but instead fixed implicitly through the expectation
values of local conservation laws.

Our main results are as follows.
\begin{enumerate}[label*=(\roman*)]
\item 
We have presented a method for calculating the expectation values of
local conservation laws in simple initial states (of product or
matrix-product form). We obtained explicit analytic expressions for
two cases of interest:  the state with all spins aligned in the
transverse direction (Section~\ref{transverse}) and the N\'eel state
(Section~\ref{Neel}). 
\item 
We have shown that as long as the expectation values of local
conservation laws are known explicitly, it is possible to avoid having
to determine the Lagrange multipliers defining the density matrix
of the generalized Gibbs ensemble. This substantially simplifies
the calculation of short-distance correlation functions in the
stationary state (Section~\ref{s:elLag}). 
\item We showed that for ``small'' quenches, as defined in the main
text, analytic expressions for correlation functions can be
obtained. The properties of the initial 
state enter through a single function (Section~\ref{s:pert}), which we
determined analytically for quenches from the N\'eel state
(Section~\ref{s:NS}). Using the results of Ref.~[\onlinecite{wuppertal1}]
for finite-temperature correlation functions in the XXZ-chain,
we obtained expressions for short distance spin correlation
functions. In particular, we obtained an analytic expression for
short-range correlators after a quantum quench from the N\'eel state
to an XXZ Hamiltonian with sufficiently large anisotropy parameter 
$\Delta$ (Section~\ref{ss:NScorr}). 
The next orders of the perturbation expansion in $1/\Delta^2$ are easily accessible as well.
\end{enumerate}

Much
work remains. For other initial states such as $|x,\uparrow\rangle$
considered in section \ref{transverse}, we need to solve the nonlinear
integral equations numerically or implement an iterative analysis
similar to that for the N\'eel state. Work on this is in progress. For 
quenches to the critical XXZ chain $|\Delta|\leq 1$ the role of the
conservation laws discovered recently \cite{nonlocal} needs to be
clarified. \\

While this work was being written up, a preprint by B. Pozsgay
appeared on the arXiv \cite{P:2013}, where the same problem is
studied by means of a quantum transfer matrix formulation, that allows
for the calculation of short-range correlators by using the
corresponding finite-temperature results. 
The main differences to our work are the treatments of
the initial state and of the Lagrange multipiers. 

\acknowledgments
We thank F. Smirnov for drawing our attention to the quantum transfer
matrix approach to GGE expectation values. This work was supported by
the EPSRC under grants EP/I032487/1 and EP/J014885/1.

\end{document}